\pdfoutput=1
\documentclass{article}
\usepackage{spconfa4,amsmath,graphicx}
\usepackage{booktabs}

\usepackage{tikz}
\usetikzlibrary{calc}
\usetikzlibrary{arrows,shapes,positioning}
\usepackage[utf8]{inputenc}
\usepackage{pgfplots}
\DeclareUnicodeCharacter{2212}{−}
\usepgfplotslibrary{groupplots,dateplot}
\usetikzlibrary{patterns,shapes.arrows}
\pgfplotsset{compat=newest}
\usepackage{amsmath}
\usepackage{subcaption}
\usepackage{paralist}
\usepackage{siunitx}
\usepackage{bbding}

\usepackage{todonotes}

\newcommand{\meanstd}[2]{\num[round-mode=places, round-precision=1]{#1} {(\footnotesize {$\pm$\,\num[round-mode=places, round-precision=1]{#2}}})}

\usepackage{xcolor}

\newcommand*{\fa}{\ensuremath{T_{fa}}}
\newcommand*{\ms}{\ensuremath{T_{miss}}}
\newcommand*{\se}{\ensuremath{T_{conf}}}
\newcommand*{\ts}{\ensuremath{T_{ref}}}

\title{Audio Diarization: A New Paradigm for Exploring Audio Recordings with Unknown Event Classes}
\name{Alexander Werning, Reinhold Haeb-Umbach}%
\address{Paderborn University, Communications Engineering Department, Germany\\ \small\ttfamily \{werning,haeb\}@nt.upb.de } %
\begin{document}
\ninept
\maketitle
\begin{abstract}

We propose a new task, audio diarization. The motivation is that there are applications, such as audio monitoring in an unknown environment, where initially the sound event classes to be recognized are unknown. For such a scenario, we propose to first localize in time relevant sound events and to classify them, e.g., by comparing with known event classes, in a second step. This contribution is dedicated to the first step, which we call audio diarization, as it is reminiscent of the speaker diarization stage that precedes and simplifies the second stage, speech recognition, in multi-talker conversational speech processing.  In this contribution, we define audio diarization as detecting onset and offset times of sound events with overlap for an open set of classes and without user prompts. We show how a speaker diarization system can be adjusted for audio diarization and propose an evaluation setup. Compared to a closed-set sound event detection system, the proposed system achieves similar performance with the additional ability to detect novel sounds.

\end{abstract}
\begin{keywords}
Diarization, Sound Event Detection, 
\end{keywords}

\section{Introduction}
Since the nascence of the detection of acoustic scenes and events (DCASE) challenge and community \cite{cornell2024dcase}, sound event detection (SED) has been one of the major research areas. SED is about the recognition of relevant, that is, pre-defined, acoustic events and localizing their occurrence in time. A common approach is to frame this problem as a binary classification over time per class \cite{mesaros2021sound}. Both classification and detection are therefore performed jointly. %

Clearly, existing SED methods are limited to detecting a predefined set of event classes. While for some applications, the insensitivity to event classes not seen in training is a desirable feature, for other applications this is a limitation. Real-world soundscapes consist of a large variety of sound classes and it is close to impossible to gather sufficient training examples %
to apply the SED paradigm to detect and classify all of them. So, if one is interested in detecting rare or new sound event classes in an environment and localizing them in time, the SED approach is inappropriate.

An extension of SED is open-vocabulary SED \cite{hai2025flexsed}, where new, unseen classes of sound events should also be classified and detected. A common approach is to view SED as a retrieval system. Instead of a fixed set of outputs per class, event detections, i.e.\ onsets and offsets, are retrieved using language prompts (e.g., ``dog barks'') 
\cite{hai2025flexsed,yin2025exploring,wu2025flam,primus2025tacos}, or audio prompts %
\cite{cai2025detect}. These systems build upon multi-modal language-audio models, such as CLAP \cite{elizalde2023clap}, to query for new event classes ad-hoc during test time. %

In open-vocabulary SED, the underlying assumption is that the user knows what to look for in order to design the prompts. But what if the user does not? When building an SED system for a new application or environment, one might not yet know the full set of event classes that is to be recognized. For example, in bioacoustics, the population of animals and thus the set of animal call types that will be recorded in a certain environment might be unknown. Detecting new call types might even be the main purpose of the whole effort. %

We propose to address the problem by separating the detection from the classification task entirely, which, to our knowledge, has not been done before. This contribution is concerned with the first, the detection task, which we call audio diarization (AD) in the following. We define it as localizing sounds in time and discriminating the detections of occurrences of the same event class from those of different event classes. The sound event classes should not need to occur in the training data or be given via a prompt for this detection. Detections are complicated by the possibility of different event classes being active at the same time. 

The proposed split into two tasks bears similarity with multi-talker speech processing, where a speaker diarization component precedes the actual classification, the speech recognition \cite{raj2021integration}. Speaker diarization is about answering the question ``who spoke when'', i.e.\ chopping the input into consistent segments with constant speaker activity and giving segments with the same speaker active the same label.  
This detection is also complicated by the occurrence of speech overlap between different speakers. Speaker diarization systems are usually designed to work with speakers not seen during training \cite{park2022review}.

We aim at transferring this ability of detecting speaker activity for unseen speakers to AD by detecting sound event activity for unseen event classes.
Current state-of-the-art systems for speaker diarization use the so-called end-to-end neural diarization (EEND) architecture %
\cite{fujita2019end, fujita2019end2}. They do not require queries or hints regarding the speakers and are therefore candidates for solving AD. %
In prior work, EEND has only been applied to SED with fixed classes \cite{jiang2025unified, cao2020event, cao2021improved}, not the AD task.
Our contribution aims at answering the following research questions:
\begin{compactitem}
    \item How can existing EEND systems for speaker diarization be adapted to perform AD?
    \item How can an AD system be evaluated?
\end{compactitem}
We propose an AD system based on an existing speaker diarization architecture and compare it to a (closed set) SED baseline. We believe this to be a fair comparison, since event open-vocabulary SED systems cannot detect new classes without an appropriate prompt, which is not given for AD. %
In order to evaluate the proposed system, we use an SED dataset and additionally mix it with audio from another dataset to simulate mixtures of seen and unseen sound event classes.

We show that the proposed system is competitive and can handle overlap between multiple event classes and also detect unseen event classes, which the baseline cannot.

\begin{figure*}[!ht]
    \centering
    \input{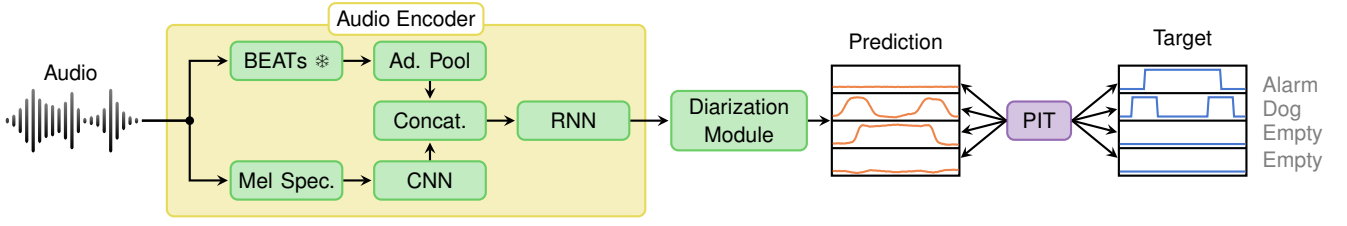}%
    \caption{Overview of the proposed audio diarization system and training procedure.}
    \label{fig:overview}
\end{figure*}
\section{Task Description and Delineation}
The task of audio diarization (AD) is introduced and the differences to existing tasks are explained in the following.
We use the term sound event to describe a single sound or a sequence of sounds originating from a sound source or class of sound sources. A single sound event can be localized in a given audio and described by a start and an end time.
For a given recording, there may be several sound events from different sources, which may overlap with each other. We are interested in detecting these sound events, i.e.\ obtaining the boundaries of these events, and grouping them, without identifying the classes of sound events via a classification into known classes.

The motivation for this task is the following: Given an audio, it may not be known which sound event classes are present, and therefore, instead of detecting only known classes, it may be useful to ask ``Where are events and which are examples of the same event class?" Once initial event boundaries are determined, they can be processed further. We hope this first processing step may allow to simplify the following tasks, similar to how speaker diarization can improve speech transcription in meeting scenarios \cite{polok2026dicow}.

The task of AD is more than a sound activity detection, where it is sufficient to determine whether any sound is active in a given audio per time frame, but no distinction between different sound classes is made and overlap between sounds is not handled. Similarly, acoustic anomaly detection is concerned with localizing anomalous sound events in time, ignoring all known or ordinary sound events. These anomalies are usually treated as a single class and overlap is not handled. For AD, we are not only interested in anomalous sounds, but all active sound events in the given audio.

For SED \cite{mesaros2021sound}, a fixed set of known event classes is defined and only events from these classes should be detected and classified. For open-vocabulary SED, this is only extended to detect events which belong to a given prompt. In contrast, for AD all events should be detected, also event classes which were not anticipated to occur by a user.

\section{An Audio Diarization System}

The proposed AD system is composed of two parts. First, an audio encoder that computes an embedding sequence for a given audio. Second, a diarization module that converts the sequence of embeddings into multiple sequences of frame-wise activity estimates. %
Figure~\ref{fig:overview} shows an overview of the system, which is explained further in the following.
\subsection{Audio Encoder}
\label{sec:audioencoder}%
The encoder module has the same architecture as the DCASE SED task 4 baseline system from 2024 \cite{cornell2024dcase}. The audio input is processed by two parallel components, a CNN and a pretrained SSL audio encoder. The CNN processes Mel spectrogram features which are extracted from the audio. For the SSL encoder, BEATs iter3 module \cite{chen2023beats} is chosen. The embedding sequence from the SSL encoder is adaptively average-pooled to have the same time resolution as the output of the CNN. Then, both SSL embeddings and CNN output are concatenated to a single frame-wise embedding sequence and processed by an RNN module to obtain frame-wise embeddings of the audio input.
We use an SSL encoder which was not fine-tuned on labeled data in order to be able to show a generalization to novel sound events without leaking sound event class information from the pre-training.

\subsection{Diarization Module}
\label{sec:diarizationmodule}
The output of the audio encoder is processed using a Transformer \cite{fujita2019end2} with two layers. The diarization module predicts frame-wise event activity probabilities for a fixed number $C$ of output channels. The order of the channels is irrelevant; it is only required that the activities within one channel belong to the same event class. The number of channels $C$ therefore constrains the maximum number of event classes that are detectable in a given audio. %

Since the order of the output channels is not relevant, no order should be enforced during training. We apply permutation-invariant training (PIT) to achieve this \cite{fujita2019end}. For each of the $K$ classes that occur in a given audio, where ${K \leq C}$, frame-wise targets are computed from the set of labeled event onsets and offsets.
For all possible assignments of these $K$ classes to the $C$ output channels, a binary cross entropy loss is computed across all time frames and channels. For channels, where no class is assigned, an empty target without activity is applied. The minimal loss over all possible assignments is then used for back-propagation. %
Although the number of possible assignments can be very large, the search for the optimal assignment can be implemented efficiently using the Hungarian algorithm.

During inference, no permutation with respect to targets needs to be resolved. The predicted event activity is obtained by applying a detection threshold of 0.5 to the frame-wise event activity probabilities for each output channel. From these binary predictions, onset and offset times can be computed for all detected events, where events in the same channel are assumed to be of the same event class.

\section{Experimental Setup}
We perform two main experiments, one to show the proposed AD system performs on par with an SED system baseline and the second to show the proposed system has the ability of detecting novel, unseen sound event classes.

\subsection{Baseline System}
The baseline system from the DCASE challenge 2024 task 4 \cite{cornell2024dcase} is compared to the proposed diarization system.
For a description of the architecture, see Section \ref{sec:audioencoder}. The BEATs encoder version was additionally fine-tuned on labeled AudioSet data, different from the proposed system, where the self-supervised checkpoint is used instead \cite{chen2023beats}, which makes the baseline pretraining not purely self-supervised, while ours is. %
The SED baseline was trained on DESED \cite{turpault2019sound} and other datasets including AudioSet-strong \cite{hershey2021benefit}. Several techniques such as output post-processing using median filters and automated hyper-parameter tuning were employed.
This presents a challenging baseline, since we omit training on multiple datasets and the other listed improvements for our system for simplicity. 

\subsection{Training Data}
The proposed system was trained only on DESED \cite{turpault2019sound}. The \textit{train\_strong}, \textit{synthetic20} and \textit{synthetic21} subsets are used as training data. Each audio clip has a length of \qty{10}{\second} and sound events are annotated via start and end times for all of the ten sound event classes in the label set. The clips are filtered such that at most four event classes occur in a single clip, allowing to set $C\!=\!4$. %
The average number of event classes per training audio is \meanstd{2.9}{1.3}, for the \textit{eval} subset it is \meanstd{1.6}{0.7}. The data is resampled to \qty{16}{\kilo\hertz}.

This dataset was curated for SED, where a distinction between relevant and irrelevant sounds is made explicitly. There are some audio recordings, where sounds, which do not belong to any of the ten labeled event types, are also present as distractors. For diarization this is unfavorable, since these events should also be detected, but we nevertheless leave it as is. %
We manually labeled several recordings of the \textit{eval} subset and found about \meanstd{30}{14.4}\% contain additional, unlabeled sound events. %

\subsection{Evaluation Metrics}

For SED systems, the Polyphonic Sound Detection Score (PSDS) is commonly reported \cite{bilen2020framework,ebbers2022threshold}. This metric presents several issues for the proposed diarization method. First, the PSDS requires a prediction for all classes of a fixed class set. In the case of AD, this can be accomplished by assigning the predictions from the output channels event classes, while defining all classes which do not occur to be inactive. This assignment will ignore the empty output channels, which are not assigned to any class. The activity on these channels is therefore ignored by the metric, although it should count as false alarms. Also, PSDS assumes that a separate detection threshold can be applied to each event class. Since the class of an output channel of the diarization system is unknown during inference, this is not possible. The PSDS is therefore not capturing the actual system performance. 

In Speaker Diarization, the commonly used metric is the diarization error rate (DER), which is the relation between the durations of false alarm time (\fa), missed speaker time (\ms) and speaker confusions (\se) to the total reference speaker time (\ts):
\begin{equation}
\text{DER} = \text{FA} + \text{MS} + \text{SC} = \frac{\fa}{\ts} + \frac{\ms}{\ts} + \frac{\se}{\ts}.
\end{equation}

The individual error rates are termed FA, MS and SC for false alarm rate, missed speaker rate and speaker confusion rate, respectively.
We keep the terminology, although speech and speaker refer to sound events and the event classes here, respectively, and apply the metric in this manner.

\subsection{Diarization of known event classes}
We compare the proposed system to the DCASE SED baseline based on the \textit{eval} subset of DESED. The results are shown in Table~\ref{tab:sed_vs_diarization_desed}. The DER of the diarization system is $3.8$ percentage points (pp) higher that the SED baseline, which we attribute to the additional training data and more complex training pipeline of the SED system. Both misses and false alarms are higher for the diarization system. %

\begin{table}[ht]
\centering
\caption{Comparison of DCASE baseline and diarization system on DESED.}
\label{tab:sed_vs_diarization_desed}
\begin{tabular}{llrrrr}
\toprule
System & DER [\%] & MS [\%] & FA [\%] & SC [\%] \\
\midrule
AD & 20.7 & 8.5 & 11.5 & 0.7 \\
SED baseline & 16.9 & 7.6 & 8.6 & 0.7 \\
\bottomrule
\end{tabular}
\end{table}
\section{Diarization of unseen event classes}
This second experiment evaluates how well sound events are detected by the diarization system when they were not seen during training. A new dataset is created to test specifically how the presence of new event classes changes the detection result.

\subsection{Dataset Creation}
\label{sec:dmix}
A second dataset was synthetically created to test whether the proposed system can handle unseen sound classes. The multi-event DESED recordings are mixed with single-event audio recordings sampled from the ESC-50 dataset \cite{piczak2015esc}.

Strong labels are not provided for the ESC-50 dataset. We use an energy-based thresholding approach to compute event on- and offsets. Specifically, the audio signals were segmented into \qty{50}{\milli\second} blocks for each of which the signal energy is computed. The blocks with an energy above -20dB compared to the maximum energy of all blocks in a given audio are labeled as active. We validated this approach by comparing it with a small portion of manually labeled audio clips.

Two datasets are created, DMix-known and DMix-unknown. For the first, six classes from ESC-50 are selected and matched to existing DESED classes: \textit{dog}, \textit{cat}, \textit{vacuum\_cleaner}, \textit{pouring\_water}, \textit{brushing\_teeth}, \textit{clock\_alarm}. The reason for testing on this first dataset with known classes is to obtain a reference error rate that reflects the increased difficulty caused by the dataset domain mismatch and by more classes and more event overlap per clip. For the second dataset, novel classes which are unrelated to the DESED classes are selected: \textit{cow}, \textit{frog}, \textit{hen}, \textit{pig}, \textit{sheep}, \textit{crow}. The classes were selected such that they do not include classes of ambient events or soundscapes without clear event onsets and offsets.

Both of these datasets are created starting from the \textit{eval} subset of DESED. For each audio clip with fewer than four active event classes, a random clip from one of the respective class sets was chosen from ESC-50 and mixed in at a randomly sampled position in time. The chosen ESC-50 audio is scaled such that the average energy matches the chosen DESED mixture. For the DMix-known dataset, special care was taken that the newly mixed-in event class is not already present in a given audio to avoid multiple simultaneous sources of the same class.  With \meanstd{2.6}{0.7}, the average number of event classes is higher than for the original DESED \textit{eval} subset.

\begin{table}[ht]
\centering
\caption{Comparison of audio diarization (AD) system and sound event detection (SED) baseline on DMix splits.}
\label{tab:diarization_vs_sed_desedmix}

\begin{tabular}{llcccc}
\toprule
System & DMix- & DER [\%] & MS [\%] & FA [\%] & SC [\%] \\
\midrule
AD  & known & \bf 35.4 & 27.6 & 4.9 & 2.9 \\
SED & known & 35.9 & 30.1 & 3.8 & 1.9 \\
\midrule
AD  & unknown & \bf 34.6 & 26.4 & 5.1 & 3.1 \\
SED  & unknown & 45.3 & 39.5 & 2.5 & 3.3 \\
\bottomrule
\end{tabular}
\end{table}

\subsection{Results}
The results of the evaluation on the DMix-known and DMix-unknown are shown in Table \ref{tab:diarization_vs_sed_desedmix}. Clearly, there is a performance gap between the SED model, which is unable to detect the unknown classes in DMix-unknown, and the diarization system. Note that DMix-unknown also contains known classes present in the original DESED clips. The DER of the SED system increases from 35.9\% DER to 45.3\% when going from DMix-known to DMix-unknown, while there is no degradation between DMix-known and DMix-unknown for the AD model, leading to a difference of almost 11pp in DER between the two systems on DMix-unknown (34.6\% vs 45.3\%). This supports the main claim of this paper, that the AD system is able to detect unseen event classes.

Comparing the results in Table~\ref{tab:sed_vs_diarization_desed} and \ref{tab:diarization_vs_sed_desedmix} we note that there is a gap of about 15pp and 19pp between the DESED and the DMix-known datasets for the systems.

To investigate what causes the gap between the DMix datasets, we evaluate the difference for specific sound events based on their source dataset. Table~\ref{tab:der_breakdown_by_source} shows the DER separately for the events sourced from DESED within DMix and those sourced from ESC-50 within DMix. The ESC-50 row represent sounds from known classes for DMix-known, and novel classes that are not in DESED for DMix-unknown. Confusion errors are excluded here, since the metrics are computed per output channel.
\begin{table}[ht]
\centering
\caption{Audio Diarization system DER breakdown by event source for the DMix datasets.}
\label{tab:der_breakdown_by_source}
\begin{tabular}{llcccc}
\toprule
DMix- & Source & DER [\%] & MS [\%] & FA [\%] \\
\midrule
known & DESED & 25.3 & 17.8 & 6.5 \\
unknown & DESED & 23.3 & 15.2 & 7.2 \\
known & ESC-50 & 35.4 & 33.5 & 1.9 \\
unknown & ESC-50 & 39.8 & 39.2 & 0.6 \\
\bottomrule
\end{tabular}
\end{table}

The gap between events sourced from DESED and those sourced from ESC-50 persists across the sets of known and novel classes. Therefore, we conclude that this is a mismatch between the dataset distributions, either the recording setups or the labeling of the events. Another source for this disparity could be the process of mixing audio from different acoustic environments.

The errors for events sourced from ESC-50 are dominated by misses, indicating that the system tends to ignore these events or predict too little activity.

\subsection{Robustness of detection threshold}
The threshold for detection was fixed to 0.5 for the previous experiments, matching prior work on speaker diarization \cite{fujita2019end,fujita2019end2}. We test the sensitivity of the systems with respect to this threshold for both the AD system and the SED baseline.

\begin{figure}
    \centering
    \begin{subfigure}[t]{0.25\textwidth}
        \centering
        \begin{tikzpicture}

\definecolor{darkgray176}{RGB}{176,176,176}
\definecolor{steelblue31119180}{RGB}{31,119,180}

\begin{axis}[
grid,
tick align=outside,
tick pos=left,
x grid style={darkgray176},
xlabel={Threshold},
xmin=-0.05, xmax=1.05,
xtick style={color=black},
y grid style={darkgray176},
ylabel={DER [\%]},
ymin=0, ymax=50,
ytick style={color=black},
scale only axis,
width=9em
]
\addplot [draw=none, draw=steelblue31119180, fill=steelblue31119180, mark=*]
table{
x  y
0.77 18.881130090255972
};
\addplot [semithick, steelblue31119180]
table {%
0 314.985212407377
0.01 68.5124739058681
0.02 49.282738042283
0.03 42.1152967745204
0.04 37.9539637068944
0.05 35.2627017651063
0.06 33.2967650139437
0.07 31.9652267806497
0.08 30.8119895662447
0.09 29.8353300692686
0.1 29.0126019808409
0.11 28.3585886910714
0.12 27.7603892984362
0.13 27.2587700939007
0.14 26.7812808378455
0.15 26.3023093825678
0.16 25.9201892283152
0.17 25.5713698379203
0.18 25.2612684370431
0.19 25.0275051086799
0.2 24.7659007756178
0.21 24.51342253845
0.22 24.2874907286293
0.23 24.0564933655368
0.24 23.8483059720929
0.25 23.6675562437735
0.26 23.4300600597648
0.27 23.267491368083
0.28 23.0984514596866
0.29 22.9523943206852
0.3 22.7922049966279
0.31 22.5476054795432
0.32 22.422661960448
0.33 22.2934608106803
0.34 22.077427709014
0.35 21.9716439880418
0.36 21.8514293433849
0.37 21.7478026281201
0.38 21.6660836923043
0.39 21.5705312147497
0.4 21.502268922366
0.41 21.4381194736686
0.42 21.3777537670277
0.43 21.3148287532294
0.44 21.1614745363951
0.45 21.1195230703933
0.46 21.0490682614374
0.47 20.8629179241113
0.48 20.7734821984043
0.49 20.7167463484081
0.5 20.6532663431419
0.51 20.4314202283102
0.52 20.3765567868844
0.53 20.3315593893183
0.54 20.2990896265926
0.55 20.2139173819185
0.56 20.153400888946
0.57 20.1339051927691
0.58 20.0545023084484
0.59 20.0147902378767
0.6 19.9715791634837
0.61 19.9468204662749
0.62 19.9489037757268
0.63 19.9165609807634
0.64 19.8501520744384
0.65 19.7971090886227
0.66 19.8047095084806
0.67 19.7717738653016
0.68 19.5115766454996
0.69 19.4931506090077
0.7 19.4520617101702
0.71 19.3774538683587
0.72 19.2322324525174
0.73 19.2138286964226
0.74 19.1378246860112
0.75 19.1550176342438
0.76 19.1878089525366
0.77 18.881130090256
0.78 18.9066192558187
0.79 18.9542735263859
0.8 18.9449824063681
0.81 18.9825351045405
0.82 19.0689706404373
0.83 19.1498634243543
0.84 19.1925428272038
0.85 19.1972939701012
0.86 19.3446194498512
0.87 19.5006349298338
0.88 19.6702663085856
0.89 19.8216443943289
0.9 19.9711038724639
0.91 20.0183818854248
0.92 20.3599695276082
0.93 20.5587916052222
0.94 21.0875287563534
0.95 21.6275533080708
0.96 22.0166376030522
0.97 22.6820427793806
0.98 24.5917197375549
0.99 27.9461533784402
1 414.6
};
\end{axis}

\end{tikzpicture}
        \caption{Audio diarization system}
    \end{subfigure}%
    \begin{subfigure}[t]{0.25\textwidth}
        \centering
        \begin{tikzpicture}

\definecolor{darkgray176}{RGB}{176,176,176}
\definecolor{steelblue31119180}{RGB}{31,119,180}

\begin{axis}[
grid,
x grid style={darkgray176},
tick align=outside,
tick pos=left,
x grid style={darkgray176},
xlabel={Threshold},
xmin=-0.05, xmax=1.05,
xtick style={color=black},
y grid style={darkgray176},
ymin=0, ymax=50,
ytick style={color=black},
scale only axis,
width=9em
]
\addplot [draw=none, draw=steelblue31119180, fill=steelblue31119180, mark=*]
table{%
x  y
0.61 16.733722580446912
};
\addplot [semithick, steelblue31119180]
table {%
0 937.412194167624
0.01 577.885255517027
0.02 313.173164497374
0.03 198.198233639727
0.04 139.567046790746
0.05 102.778793509354
0.06 79.7863675900842
0.07 64.8239141121573
0.08 54.3654061322379
0.09 47.1235465245505
0.1 41.9991596550256
0.11 37.8774509015452
0.12 34.7400477453311
0.13 32.3553338525256
0.14 30.4328412544093
0.15 28.9177851188628
0.16 27.5604583144575
0.17 26.4246697444929
0.18 25.4932389964318
0.19 24.6338378321516
0.2 23.8242829292626
0.21 23.2397488195162
0.22 22.5634736170567
0.23 21.9746498421214
0.24 21.4182350428454
0.25 20.9596360382673
0.26 20.6089291806902
0.27 20.2236802980421
0.28 19.7246305578022
0.29 19.3881529830424
0.3 19.0779721663957
0.31 18.90286665768
0.32 18.7130355941712
0.33 18.5293915386061
0.34 18.4069668601532
0.35 18.2282643535425
0.36 18.0201477053939
0.37 17.9673069355599
0.38 17.8577066296033
0.39 17.7759088659319
0.4 17.6153202603967
0.41 17.5151946507302
0.42 17.4499048396145
0.43 17.3631293418734
0.44 17.317542121789
0.45 17.0763917168741
0.46 17.0670353197996
0.47 17.0518922070267
0.48 17.0152096145947
0.49 17.0111142448437
0.5 16.9070905628063
0.51 16.9619051903336
0.52 16.9402784018585
0.53 16.9150668953875
0.54 16.9854108166209
0.55 16.8592008244973
0.56 16.8247789123153
0.57 16.8386615251483
0.58 16.8413213010991
0.59 16.8709447582758
0.6 16.9432244088332
0.61 16.7337225804469
0.62 16.7609818208099
0.63 16.8328466070645
0.64 16.8821765930889
0.65 16.9698116125913
0.66 17.0595177107923
0.67 17.165616512597
0.68 17.1328754292314
0.69 17.1844813464809
0.7 17.3344464129938
0.71 17.4215229143212
0.72 17.6108367752376
0.73 17.6953269526079
0.74 17.8903690721135
0.75 18.0316067594464
0.76 17.9135821537364
0.77 18.0082060652042
0.78 17.9959205616892
0.79 18.046582039526
0.8 18.4096663888764
0.81 18.8035657102246
0.82 19.2141605880825
0.83 19.7088617483109
0.84 20.0268491548643
0.85 20.3235674773756
0.86 20.6928006850044
0.87 20.9431659245362
0.88 21.5762808926982
0.89 22.3911974004342
0.9 23.2054073211867
0.91 24.085264539625
0.92 25.3151939969031
0.93 26.8928133190776
0.94 28.7401657148416
0.95 30.7112763856047
0.96 34.0064808926109
0.97 38.2145730501599
0.98 42.2715789911513
0.99 51.8507245894555
1 414.6
};
\end{axis}

\end{tikzpicture}
        \caption{SED baseline system}
    \end{subfigure}%
    \caption{Sensitivity of audio diarization and SED system to decision threshold value on DESED. The best DER is highlighted.}
    \label{fig:threshold}
\end{figure}
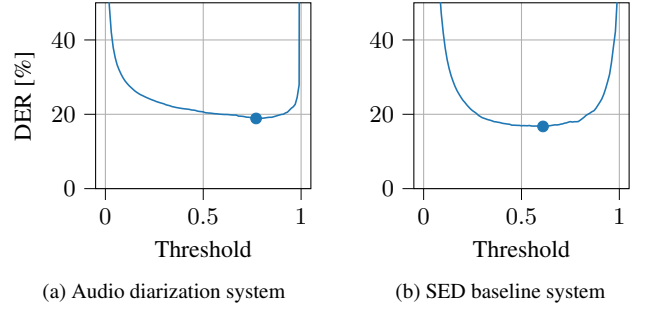

The results are shown in Figure~\ref{fig:threshold}. Both systems are not very sensitive to the threshold value, forming a wide flat section between 0.3 and 0.7. The results for the diarization system are skewed, with the best threshold at 0.77. This tendency towards higher thresholds might be due to fewer false alarms from detections of unlabeled events within DESED,  which the SED system successfully ignores.

\subsection{Event class re-identification}
The AD system is not only supposed to detect unseen event classes, but all events of a new class should be placed on the same output channel of the model. We test this explicitly by creating a new dataset, DMultiMix, created similarly to DMix described in Sec.~\ref{sec:dmix}, but instead of mixing in a single ESC-50 audio clip, we choose two of the same class and place them in the first and second half of the DESED audio clip, respectively. This is done for the known and unknown class sets.

The evaluation considers the possibility of both mixed-in events being assigned to the same output channel (Same), one event (Single Miss) or both events missing (All Miss) or them being assigned to different channels (Split). The best assignment of output channels to event targets is found as described in Sec.~\ref{sec:diarizationmodule} for each of the error cases, of which the one with the lowest loss value is chosen.

\begin{table}[ht]
    \centering
    \caption{Evaluation of errors by the Audio Diarization system by type  for the DMultiMix datasets, all values in \%.}
    \label{tab:multimix}
    \begin{tabular}{lccccc}
        \toprule
        DMultiMix- & Same & Single Miss & Both Miss & Split \\
        \midrule
        known & 26.7 & 18.4 & 53.9 & 1.0\\
        unknown & 26.2 & 17.4 & 55.8 & 0.6 \\
        \bottomrule
    \end{tabular}
\end{table}
The results are presented in Table~\ref{tab:multimix}. There is no obvious degradation from the known to the unknown set. For more than half of the audio clips, both mixed-in events are missed. A miss of one of the events also occurs frequently, which is both consistent with the finding in Table~\ref{tab:der_breakdown_by_source}, indicating the domain gap could be an issue. Further, the case of splitting the two new events is very rare.

\section{Conclusion}
A novel task for audio processing, audio diarization, was presented and a system proposed to solve this task. The proposed system is composed of an audio encoder and a diarization module, which are trained in the same fashion as a EEND system for speaker diarization. Through experiments the system was validated against an SED system as a baseline. The proposed system is able to handle novel classes of sound events which were not seen in model training. The experiments were exploratory in nature, no large audio databases were used and the diarization model is small in scale.
Future work includes the evaluation on larger datasets and the establishment of benchmark datasets better suited to the diarization task.

\bibliographystyle{IEEEbib}
\bibliography{refs}

\end{document}